# Analytical Method for Metasurface-Based Cloaking Under Arbitrary Oblique Illumination

Yi Zhang[1,2], Haiyan Fan[3], Yujie Zhang[2], Lixin Ran[1], Dexin Ye[1], and Xudong Chen[2,*]

[1] *Laboratory of Applied Research on Electromagnetics, Zhejiang University, Hangzhou 310027, China*
[2] *Department of Electrical and Computer Engineering, National University of Singapore, Singapore 117583, Singapore*
[3] *Department of Mechanical Engineering, The Hong Kong Polytechnic University, Hung Hom, Kowloon, Hong Kong SAR, China*
[*] *elechenx@nus.edu.sg*

**Abstract:** The performance of antennas can severely deteriorate in the presence of adjacent electrically-large scatterers. In this work, we use a conducting hollow cylinder to shield the scatterer. The cylinder is shelled with single layer dielectric and electromagnetic metasurface. The scattering field analysis with respect to the surface impedance is derived. By optimizing the anisotropic impedance distribution, the scattering cross-section can be effectively reduced. The proposed method is valid for both TM$^z$, TE$^z$ and non-TM/TE incident field. The accuracy and effectiveness of the method are verified by four cloaking scenarios in microwave regime. We demonstrate that with the surface impedance obtained by our method, a metasurface is designed with physical subwavelength structure. We also show a cloaking scenario under magnetic dipole radiation, which is closer to the case of a realistic antenna. This method can be further applied to cloaking tasks in terahertz and optical regimes.

## 1. Introduction

With the arrival of 5G era, the demand on wireless data transmission service is sharply increasing. To achieve a full-coverage of 5G service, a growing number of base stations in urban and rural area are set up like mushrooms after rain. It is often the case that some unexpected scatterer neighbors exist near the base stations, blocking the radiation of the antennas or at least interfering antennas' beamform and, consequently, degrading the overall quality of wireless data transmission. 5G communication system also provides generous capacity for the wireless sensor network. In the indoor environment, diverse miniaturized sensors form more intelligent lifestyles. However, the wireless sensors can be easily influenced by the nearby furniture, appliance, etc., resulting in poor electromagnetic compatibility. Some antennas work well on their own, but are necessarily equipped with supporting structure which can be a significant scatterer due to an extremely close distance to the antennas.

To address this problem, prior works aimed at making the scatterer invisible [1, 2]. A well-known technology is transformation optics (TO) [3], which can achieve perfect cloaking regardless of the shape of scatterer and the incident field, and it was physically realized in microwave regime [4] and can be also successfully applied to carpet cloaking [5-7]. However, the most challenging part of TO is the highly anisotropic properties of the metamaterial both in permittivity and permeability.

Another path is scattering cancellation technology [8]. The main idea of scattering cancellation is to suppress the dominant term of scattering harmonics. As an effective method, the plasmonic or mantle cloaking is proposed, in which the scatterer is conformally enclosed by single plasmonic layer [9, 10] or stacked layers [11, 12]. These additional layers can produce polarizability opposite to that produced by the scatterer. To physically realize the desired covering, the metamaterial is used to generate negative polarizability, for instance, the parallel-plate waveguide working under cutoff frequency [13-15], the mixture of dielectric and metal [16, 17], and metal-plate structure [18]. Other than fully-cover fashion, the satellite antiphase scatterers [19] or surrounding nanoparticles [20, 21] are also effective to reduce the total scattering, compared to mantle cloaking, the introducing of satellite antiphase scatterers provides a noninvasive way to cloak an object. The scattering cancellation is also applied to the design of wideband reflectionless electromagnetic windows [22, 23] to reduce the profile of antennas [24].

Electromagnetic metasurface-based cloaking has gained increasing attention over the past decades. As a 2D counterpart of 3D metamaterial, the metasurface is with subwavelength thickness, manipulates the behaviors of electromagnetic waves through specific boundary condition rather than volumetric constitutive parameters. The metasurfaces based on generalized Snell's Law [25] are designed for carpet cloaking [26, 27], freestanding cloaking [28] and beam controlling [29-33].

In the meanwhile, Huygens' surface is also substantially developed [34-37]. Compared with the electromagnetic metasurface, the magnetic current can also be induced on the Huygens' surface, leading to the discontinuity of electric field. The existences of magnetic currents provide additional degrees of freedom for reconstruction and manipulation of the electromagnetic field. Active metasurface [38] allows the negative resistance of the surface impedance. The active part acts as the secondary source induced by the incident field, and re-radiates the augmented field to compensate the shadow area that blocked by the scatterer [39-41]. However, the realization of negative resistance metasurface involved additional microwave circuits and antennas, increasing the complexity of the total system.

D.H. Kwon did a series of works on lossless and passive metasurface cloaking with scalar [42] and tensorial surface impedance [43, 44]. The designed anisotropic surface impedance is spatially modulated and is able to cloak large size PEC cylinder under the normal illumination of $TM^z$ wave. These works significantly reduce the difficulty of physical implementation using active metasurface. Similar works but with isotropic surface impedance can be also found in [45]. The isotropic impedance surface is physically realized with Jerusalem cross structure. [46] presents a concentric structure involving multilayer anisotropic metasurfaces. Since the innermost layer is nonreciprocal, the whole system inherits the nonreciprocity. Such design can be applied in stealth, blockage avoidance, illusion and cooling. In realistic application scenarios, the incident field cannot always be ideally normal to the scatterer. For cloaking under oblique incident field, [47] analyzes the distribution of all field components under $TM^z$ oblique incident field, and an optimal isotropic surface impedance is obtained to reduce the scattering cross-section (SCS). Note that the radius of the cloaked cylinder is in the range of quasi-static limit ($\lambda/20$), which means only the first order harmonics is dominant and needs to be cancelled. [12] proposed a co-located dual-band antenna scheme, to make the low-frequency (LF) antennas' arms transparent to the high-frequency (HF) antennas, and the arms are shelled with multilayer isotropic dielectrics. Under oblique plane wave illumination, the SCS of the LF antennas' arms is effectively reduced and the radiation pattern of the HF antenna is preserved. Also, the cross-section of arm is also in the range of quasi-static limit.

In this work, we propose a method to cloak an electrically-large cylindrical scatterer based on an anisotropic lossless and passive metasurface. This method is applicable for arbitrary 2D or 2.5D ($z$-harmonic) incident wave illumination. We mathematically derive the scattering field dependency on the incident field, surface impedance, thickness and relative permittivity $\varepsilon_r$ of the host medium. With a fixed $\varepsilon_r$, the thickness of the host medium and the surface impedance can be optimized to achieve low-scattering performance. Four full-wave simulated examples are presented, including cloaking an electrically-large conducting cylinder under $TE^z$ normal, $TE^z$ oblique, $TM^z$ normal, and $TM^z$ oblique illumination, respectively, to verify the accuracy and the effectiveness of our method. Particularly, a physical realization of an impedance metasurface is designed for cloaking under $TE^z$ wave normal illumination. The cloaking of a finite-length cylinder under magnetic dipole radiation proves the potential for realistic radiation cases, for instance, cloaking the scatterer illuminated by an actual antenna.

## 2. Problem Formulation

Near the antenna, the presence of various of scatterer will interfere the radiation. In the most cases, there is no analytical solution of the scattering field for arbitrary shape of objects in Maxwell's equations. The numerical solution for cloaking arbitrary shape is very complicated and always require massive computational resource. In fact, the numerical solution is also unnecessary – the customized conformal covering is no longer suitable for another different object, indicating poor reusability. To simplify this problem, an analyzable geometry in Maxwell's equation can be chosen to cover the object under cloaking, for instance, a cylindrical cover made of metal with negligible loss. The interior is totally shielded and has no influence on external EM field. With the regular shape, the exact solution can be derived.

At this stage, the problem has been converted to cloaking a cylindrical conducting core. The thin electromagnetic metasurface is adopted for realization. The EM metasurface is designed with varying surface impedance, which can be locally defined as

$$E(r) = [Zs(r)] \cdot \hat{n} \times (H_1(r) - H_2(r)), \quad (1)$$

where Zs is the surface impedance which can be either a tensor or a scalar. $\hat{n} \times (H_1(r) - H_2(r))$ is the discontinuity of the tangential magnetic field (H-field) across the EM metasurface, which can be regarded as a surface electric current. Across the EM metasurface interface, the continuity of electric field (E-field) is maintained.

The direct contact between conducting core and metasurface can lead to a short circuit. To avoid this, the dielectric layer is used to cover the conducting core. It also acts as the host medium on which the metal can be patterned to characterize the surface impedance.

As shown in Fig. 1, the innermost is a cylindrical conducting core with a radius of $r_1$. The size is chosen to completely encompass the object under cloaking. Outside the core is the conformal dielectric cover ring with outer radius of $r_2$ and relative permittivity of $\varepsilon_r$, this host medium is concentric with the conducting core. The impedance surface is adhered on the outermost surface of the host medium acting as the interface between the host medium and the free space.

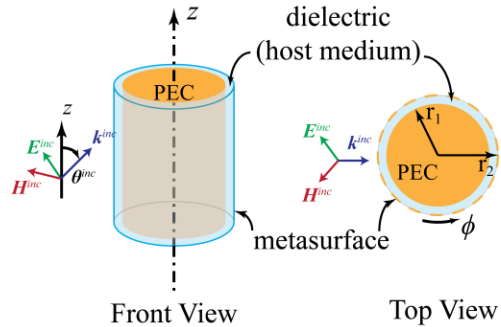

Fig. 1. The schematic of the cloaked cylinder under the illumination of an oblique incident field.

In our work, the $\rho$-$\phi$-$z$-cylindrical coordinate is used, as shown in Fig. 1. The scattering field will not be depolarized (i.e., producing cross-polarized scattering field) under normal illumination, or solely conducting cylinder exists without dielectric cover [48, 49]. Since the host medium is indispensable, the cross-polarization should be taken into consideration for oblique incidence cloaking. To start with, the $\rho$ and $\phi$ components of E-field and H-field can be expressed as follows:

$$\begin{cases} E_\phi = \left( \dfrac{k_z}{jk_0^2\rho} \dfrac{\partial E_z}{\partial \phi} - \dfrac{1}{j\omega\varepsilon} \dfrac{\partial H_z}{\partial \rho} \right) \dfrac{k_0^2}{k_t^2} \\[4pt] E_\rho = \left( \dfrac{1}{j\omega\varepsilon\rho} \dfrac{\partial H_z}{\partial \phi} - \dfrac{jk_z}{k_0^2} \dfrac{\partial E_z}{\partial \rho} \right) \dfrac{k_0^2}{k_t^2} \\[4pt] H_\phi = \left( \dfrac{1}{j\omega\mu} \dfrac{\partial E_z}{\partial \rho} + \dfrac{k_z}{jk_0^2\rho} \dfrac{\partial H_z}{\partial \phi} \right) \dfrac{k_0^2}{k_t^2} \\[4pt] H_\rho = \left( -\dfrac{1}{j\omega\mu\rho} \dfrac{\partial E_z}{\partial \phi} - \dfrac{jk_z}{k_0^2} \dfrac{\partial H_z}{\partial \rho} \right) \dfrac{k_0^2}{k_t^2} \end{cases}, \quad (2)$$

which are dependent on the *z*-components of E-/H-field (E*z*/H*z*-field) in cylindrical coordinate. For any 2-D incident wave, the *z*-component can be written in the superposition of every order of cylindrical harmonics:

$$\begin{cases} E_z^{inc} = \sum_{n=-\infty}^{\infty} a_n^e J_n(k_t\rho) e^{jn\phi} \\ H_z^{inc} = \sum_{n=-\infty}^{\infty} a_n^h J_n(k_t\rho) e^{jn\phi} \end{cases}, \quad (3)$$

where $a_n^e$ and $a_n^h$ are the coefficients of incident harmonics for E*z*-field and H*z*-field, which can be obtained by cylindrical wave expansion, $k_t$ and $k_z$ are the transverse wavenumber and longitudinal wavenumber in the free space, respectively. The harmonics term along *z*-dimension exp(-*jk*$_z$*z*) is omitted. The Bessel function of the first kind $J_n(\cdot)$ stands for the cylindrical standing wave. The scattering E*z*-/H*z*-field in the free space can be also expanded in cylindrical harmonics superposition and parameterized as

$$\begin{cases} E_z^{sca} = \sum_n b_n^e H_n^{(2)}(k_t\rho) e^{jn\phi} \\ H_z^{sca} = \sum_n b_n^h H_n^{(2)}(k_t\rho) e^{jn\phi} \end{cases}, \quad (4)$$

where $b_n^e$ and $b_n^h$ are the coefficients of scattering harmonics for E*z*-/H*z* field, $H_n^{(2)}(\cdot)$ is the Hankel function of the second kind, standing for cylindrical outgoing wave in *j*-convention. While, Hankel function of the first kind $H_n^{(1)}(\cdot)$ stands for cylindrical ingoing wave. The E*z*-/H*z*-field inside the dielectric host medium (internal field) can be parameterized as the superpositions of both ingoing and outgoing waves

$$\begin{cases} E_z^{int} = \sum_n \left( c_n^e H_n^{(1)}(k_{rt}\rho) + d_n^e H_n^{(2)}(k_{rt}\rho) \right) e^{jn\phi} \\ H_z^{int} = \sum_n \left( c_n^h H_n^{(1)}(k_{rt}\rho) + d_n^h H_n^{(2)}(k_{rt}\rho) \right) e^{jn\phi} \end{cases}, \quad (5)$$

where $k_{rt}$ is the transverse wavenumber in the dielectric. According to Eq. (2), E$\phi$-field and H$\phi$-field across the boundary can be calculated as

$$\begin{cases} E_\phi^{out} = \dfrac{k_0^2}{k_t^2} \left( \begin{array}{l} \dfrac{k_z}{jk_0^2\rho} \sum_n jn\left( a_n^e J_n(k_t\rho) + b_n^e H_n^{(2)}(k_t\rho) \right) e^{jn\phi} \\ -\dfrac{1}{j\omega\varepsilon_0} k_t \sum_n \left( a_n^h J_n'(k_t\rho) + b_n^h H_n^{(2)\prime}(k_t\rho) \right) e^{jn\phi} \end{array} \right) \\[20pt] E_\phi^{int} = \dfrac{k_d^2}{k_{dt}^2} \left( \begin{array}{l} \dfrac{k_z}{jk_d^2\rho} \sum_n jn\left( c_n^e H_n^{(1)}(k_{dt}\rho) + d_n^e H_n^{(2)}(k_{dt}\rho) \right) e^{jn\phi} \\ -\dfrac{1}{j\omega\varepsilon_0\varepsilon_r} k_{dt} \sum_n \left( c_n^h H_n^{(1)\prime}(k_{dt}\rho) + d_n^h H_n^{(2)\prime}(k_{dt}\rho) \right) e^{jn\phi} \end{array} \right) \end{cases} \quad (6)$$

$$\begin{cases} H_\phi^{out} = \dfrac{k_0^2}{k_t^2}\left( \dfrac{k_z}{jk_0^2\rho}\sum_n jn\left(a_n^h J_n(k_t\rho)+b_n^h H_n^{(2)}(k_t\rho)\right)e^{jn\phi} \\ +\dfrac{1}{j\omega\mu_0}k_t\sum_n \left(a_n^e J_n{}'(k_t\rho)+b_n^e H_n^{(2)}{}'(k_t\rho)\right)e^{jn\phi} \right) \\[2mm] H_\phi^{int} = \dfrac{k_d^2}{k_{dt}^2}\left( \dfrac{k_z}{jk_d^2\rho}\sum_n jn\left(c_n^h H_n^{(1)}(k_{dt}\rho)+d_n^h H_n^{(2)}(k_{dt}\rho)\right)e^{jn\phi} \\ +\dfrac{1}{j\omega\mu_0}k_{dt}\sum_n \left(c_n^e H_n^{(1)}{}'(k_{dt}\rho)+d_n^e H_n^{(2)}{}'(k_{dt}\rho)\right)e^{jn\phi} \right) \end{cases},$$

in which the prime acts on the whole variable in the bracket. The freedom of unknowns can be reduced by matching boundary conditions. Firstly, at the surface of the conducting core $\rho=r_1$, $Ez$-field and $E\phi$-field should be zero, so that the relation of $d_n^{e(h)}$ and $c_n^{e(h)}$ can be obtained as

$$\begin{cases} d_n^e = -c_n^e H_n^{(1)}(k_{rt}r_1)/H_n^{(2)}(k_{rt}r_1) \triangleq f_n^e c_n^e \\ d_n^h = -c_n^h H_n^{(1)}{}'(k_{rt}r_1)/H_n^{(2)}{}'(k_{rt}r_1) \triangleq f_n^h c_n^h \end{cases}. \tag{7}$$

Secondly, the continuity of $Ez$-field at the dielectric-free space interface gives

$$a_n^e J_n(k_t r_2)+b_n^e H_n^{(2)}(k_t r_2) = c_n^e\left[H_n^{(2)}(k_{rt}r_2)+f_n^e H_n^{(2)}(k_{rt}r_2)\right]$$
$$\Rightarrow c_n^e \triangleq g_{an}^e a_n^e + g_{bn}^e b_n^e \tag{8}$$

The equation for the continuity of $E\phi$-field at the interface is a little complicated, but is still derivable, which takes the form of

$$k_t c_n^h\left(H_n^{(1)}{}'(k_{dt}r_2)+f_n^h H_n^{(2)}{}'(k_{dt}r_2)\right) - k_{dt}\left(a_n^h J_n{}'(k_t r_2)+b_n^h H_n^{(2)}{}'(k_t r_2)\right)$$
$$= \dfrac{\omega\varepsilon_0 k_z}{r_2 k_t k_{dt}}(\varepsilon_r - 1)\sum_n (jn)\left(a_n^e J_n(k_t r_2)+b_n^e H_n^{(2)}(k_t r_2)\right)e^{jn\phi} \tag{9}$$
$$\Rightarrow c_n^h \triangleq g_{an}^h a_n^h + g_{bn}^h b_n^h + g_{an}^{he} a_n^e + g_{bn}^{he} b_n^e$$

where $g$ coefficients are the functions of Bessel functions and Hankel functions. The difference of tangential H-field across the metasurface boundary can be rewritten with respect to $a_n^{e(h)}$ and $b_n^{e(h)}$

$$\begin{aligned} H_\phi^{diff}\Big|_{\rho=r_2} &= \left[H_\phi^{out}-H_\phi^{int}\right]_{\rho=r_2} \triangleq \sum_{n=-\infty}^{\infty}\left(S_{a,n}^{\phi,e}a_n^e+S_{a,n}^{\phi,h}a_n^h+S_{b,n}^{\phi,e}b_n^e+S_{b,n}^{\phi,h}b_n^h\right)e^{jn\phi} \\ H_z^{diff}\Big|_{\rho=r_2} &= \left[H_z^{out}-H_z^{int}\right]_{\rho=r_2} \triangleq \sum_{n=-\infty}^{\infty}\left(S_{a,n}^{z,e}a_n^e+S_{a,n}^{z,h}a_n^h+S_{b,n}^{z,e}b_n^e+S_{b,n}^{z,h}b_n^h\right)e^{jn\phi} \end{aligned}, \tag{10}$$

and the tangential E-field at the metasurface boundary can be expressed with respect to $a_n^{e(h)}$ and $b_n^{e(h)}$

$$\begin{aligned} E_\phi\Big|_{\rho=r_2} &\triangleq \sum_{n=-\infty}^{\infty}\left(T_{a,n}^{\phi,e}a_n^e+T_{a,n}^{\phi,h}a_n^h+T_{b,n}^{\phi,e}b_n^e+T_{b,n}^{\phi,h}b_n^h\right)e^{jn\phi} \\ E_z\Big|_{\rho=r_2} &\triangleq \sum_{n=-\infty}^{\infty}\left(T_{a,n}^{z,e}a_n^e+T_{b,n}^{z,e}b_n^e\right)e^{jn\phi} \end{aligned}, \tag{11}$$

where $S$ and $T$ coefficients are the functions of Bessel functions and Hankel functions, related to the quantities of H-field and E-field, respectively. According to Eq. (1), the relationship between tangential field and surface impedance is

$$\begin{cases} E_z(\phi) = Zs(\phi)H_\phi^{diff}(\phi) \\ E_\phi(\phi) = -Zs(\phi)H_z^{diff}(\phi) \end{cases}, \tag{12}$$

where the surface impedance $Zs$ is assumed to be a scalar function of $\phi$ and is locally isotropic. In comparison, [43] used bianisotropic tensorial surface impedance. Expanding Eq. (12),

$$\begin{cases} Zs(\phi) = \dfrac{\sum_{n=-\infty}^{\infty}\left(T_{a,n}^{z,e}a_n^e + T_{b,n}^{z,e}b_n^e\right)e^{jn\phi}}{\sum_{n=-\infty}^{\infty}\left(S_{a,n}^{\phi,e}a_n^e + S_{a,n}^{\phi,h}a_n^h + S_{b,n}^{\phi,e}b_n^e + S_{b,n}^{\phi,h}b_n^h\right)e^{jn\phi}} \\ Zs(\phi) = \dfrac{\sum_{n=-\infty}^{\infty}\left(T_{a,n}^{\phi,e}a_n^e + T_{a,n}^{\phi,h}a_n^h + T_{b,n}^{\phi,e}b_n^e + T_{b,n}^{\phi,h}b_n^h\right)e^{jn\phi}}{\sum_{n=-\infty}^{\infty}\left(S_{a,n}^{z,e}a_n^e + S_{a,n}^{z,h}a_n^h + S_{b,n}^{z,e}b_n^e + S_{b,n}^{z,h}b_n^h\right)e^{jn\phi}} \end{cases}. \tag{13}$$

It can be seen both the denominators and numerators are in Fourier series style, therefore the surface impedance can be also written in Fourier series style

$$Zs(\phi) = \sum_{n=-\infty}^{\infty} z_n \exp(jn\phi), \tag{14}$$

in which $z_n$ is the coefficient of the $n$-th term of Fourier series. If $Zs(\phi)$ is a constant (i.e., for $n \neq 0$, $z_n = 0$), there will be no interactions between different order $n$ and the metasurface become isotropic. The interactions between different orders of harmonics can be characterized by anisotropic metasurface in the matrices form

$$\begin{cases} [T_a^{z,e}][\boldsymbol{a^e}]+[T_b^{z,e}][\boldsymbol{b^e}] = Zs\begin{pmatrix} [S_a^{\phi,e}][\boldsymbol{a^e}]+[S_b^{\phi,e}][\boldsymbol{b^e}] \\ +[S_a^{\phi,h}][\boldsymbol{a^h}]+[S_b^{\phi,h}][\boldsymbol{b^h}] \end{pmatrix} \\ \begin{pmatrix} [T_a^{\phi,e}][\boldsymbol{a^e}]+[T_b^{\phi,e}][\boldsymbol{b^e}] \\ +[T_a^{\phi,h}][\boldsymbol{a^h}]+[T_b^{\phi,h}][\boldsymbol{b^h}] \end{pmatrix} = -Zs\begin{pmatrix} [S_a^{\phi,e}][\boldsymbol{a^e}]+[S_b^{\phi,e}][\boldsymbol{b^e}] \\ +[S_a^{\phi,h}][\boldsymbol{a^h}]+[S_b^{\phi,h}][\boldsymbol{b^h}] \end{pmatrix} \end{cases}, \tag{15}$$

where $\boldsymbol{T_a^{z,e}} = \text{diag}[T_{a,n}^{z,e}]$, and so on, being the diagonal matrix spanned by the coefficients of $a_n^{e(h)}$ and $b_n^{e(h)}$ in the numerator and denominator in Eq. (13). The column vectors $\boldsymbol{a_n^{e(h)}} = \text{vec}[a_n^{e(h)}]$ and $\boldsymbol{b_n^{e(h)}} = \text{vec}[b_n^{e(h)}]$.

$$\boldsymbol{Zs} = \begin{bmatrix} \ddots & & & & \\ & z_0 & z_{-1} & z_{-2} & \\ & z_1 & z_0 & z_{-1} & \\ & z_2 & z_1 & z_0 & \\ & & & & \ddots \end{bmatrix} \tag{16}$$

being a Toeplitz matrix. A straightforward interpretation of $Zs$ matrix is the relationship for harmonics order conversion, for instance, $z_0$ is the conversion weight between the same order harmonics of tangential E-field and H-field, and $z_{-1}$ is the conversion weight between the $n$-th order harmonics of H-field and the $(n-1)$-th order harmonics of E-field, etc. Now, we are able to calculate the scattering field coefficients $b_n^{e(h)}$ of each order with respect to the incident field coefficients $a_n^{e(h)}$ by solving matrices equation Eq. (15). As for the cloaking purpose, our objective function is defined as

$$y = \underset{\cdots z_{-1}, z_0, z_1, \cdots}{\arg\min} \sum_{n=-\infty}^{\infty} \left|b_n^e\right|^2 + \left|b_n^h\right|^2. \tag{17}$$

By minimizing all the coefficients of scattering $Ez$-/$Hz$-field, the total scattering field can be reduced. Note that, cloaking under TM or TE incident field is the special case, for which $a_n^h$ or $a_n^e$ are zero. It can be further simplified when the cylindrical scatterer is normally illuminated by eliminating the terms in Eq. (6) that include $k_z$, as $k_z = 0$.

### 3. Numerical Calculations and Simulations

Based on the above analysis and derivation, we first provide an example of a cloaking scenario under 10GHz TE wave illumination – the incident magnetic field is parallel to $z$-axis, while the incident electric field is $y$-polarized. The radius of the inner conducting core is one wavelength in the free space ($\lambda_0$), and the relative permittivity of the dielectric shell is 20.

In the optimization procedure, the cylindrical harmonics order of the incident wave is truncated from $n = -50\sim 50$, i.e., 101 orders in total, which can sufficiently approximate the plane wave illumination. For the coefficients $z_n$ of the Fourier expansion of $Zs(\phi)$, $n$ is selected from -10 to 10, and to achieve a purely imaginary surface impedance value – a lossless and passive one, $z_n$ should be anti-conjugate symmetric: assuming the complex number $z_n = r_n + j \cdot x_n$, one should satisfy $r_n = - r_{-n}$ and $x_n = x_{-n}$. In this case, the number of optimizable parameters is reduced to nearly a half. Next, considering the geometric symmetry and incident wave symmetry about $xOz$ plane, the impedance distribution can be also symmetric about $xOz$ plane. Thus, for any $n$, $r_n = 0$. Therefore, the number of optimizable parameters is further reduced. The optimized impedance Fourier coefficients and the optimized outer radius of the host medium are given in Table 1.

Table 1. The Optimizable Parameters in TE Normal Cloaking Scenario ($\theta_{in} = 90°$)

| Parameter | $z_0$ | $z_{\pm 1}$ | $z_{\pm 2}$ | $z_{\pm 3}$ | $z_{\pm 4}$ | $z_{\pm 5}$ |
|---|---|---|---|---|---|---|
| value | -18.63 | -0.03815 | -0.6824 | 0.01339 | 0.003739 | -0.03307 |
| Parameter | $z_{\pm 6}$ | $z_{\pm 7}$ | $z_{\pm 8}$ | $z_{\pm 9}$ | $z_{\pm 10}$ | $r_2$ |
| value | 0.3627 | 0.05407 | 1.162 | -0.1302 | 2.019 | $1.01\lambda_0$ |

According to Eq.x, the surface reactance value around the circumference is plotted in Fig. 2(a). The surface resistance is zero and thus is omitted here. We place the field probe circle at the radius $\rho_{probe} = 8\lambda_0$, which can be regarded in the far-field region. The $z$-component of scattering H-field at the probe circle is plotted in Fig. 2(b).

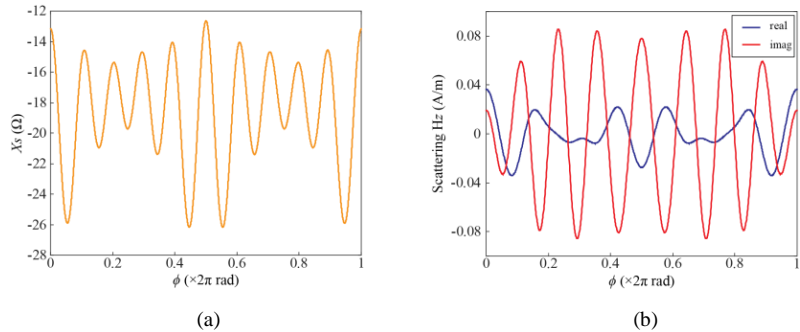

(a)                                    (b)

Fig. 2. (a) The optimized surface reactance value around the outer circumference of the host medium. The impedance is isotropic along $z$-axis. (b) The calculated $z$-component of scattering H-field (real part and imaginary part) at the field probe circle $\rho_{probe} = 8\lambda_0$.

To verify the accuracy and the effectiveness of our calculation, we implement the full-wave simulation in a commercial software Comsol Multiphysics®. Fig. 3(a) shows $z$-component of scattering H-field at the same probe circle. It can be seen that the simulated field agrees well with our calculated result in Fig. 2(b). The scattering cross-section (SCS) of the cloaked conducting cylinder is shown in Fig. 3(b). For comparison, the SCS of a bare conducting cylinder with the same size as the conducting core of the cloaked cylinder is also plotted in Fig. 3(b). It can be seen that the SCS reduces by nearly 10dB on average for all directions. Especially, the forward scattering (0°) is greatly reduced from -10dB to -35dB.

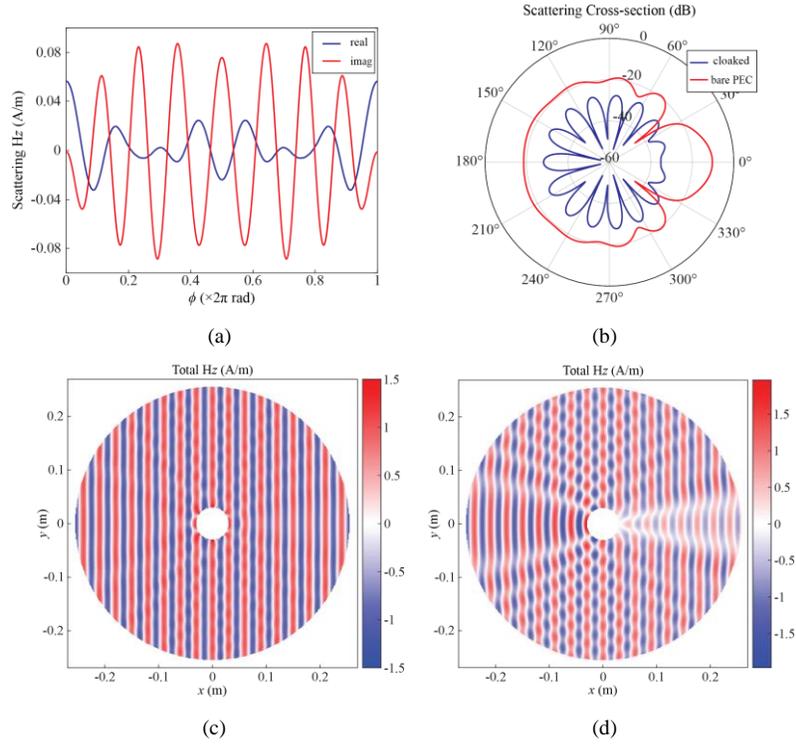

Fig. 3. (a) The simulated $z$-component of scattering H-field (real part and imaginary part) at the field probe circle $\rho_{probe} = 8\lambda_0$. (b) The scattering cross-sections of a cloaked conducting cylinder and the bare conducting core. (c) The total H-field of the cloaked conducting cylinder. (d) The total H-field of the bare conducting core.

To further investigate the performance, the total H-fields in the presence of the cloaked cylinder and the bare one are shown in Fig. 3(c) and Fig. 3(d), respectively. It can be seen that in the presence of the cloaked cylinder, most of the incident field is maintained.

For another example of cloaking under oblique wave illumination, the incident angle $\theta_{in}$ is set to be $60°$. The magnetic incident field is parallel to $y$-axis. And the included angle between electric incident field and $z$-axis is $30°$. The radius of the inner conducting core and the relative permittivity of the dielectric shell are unchanged. Table 2 lists the optimized impedance Fourier coefficients and the outer radius of the host medium. The calculated surface reactance distribution along the circumference is presented in Fig. 4(a), and the comparison of the SCSs of the cloak conducting cylinder and the bare conducting core is shown in Fig. 4(b). It can be observed that scattering field in the most angles are reduced, especially for the forward scattering field.

**Table 2. The Optimizable Parameters in TM Oblique Cloaking Scenario ($\theta_{in} = 60°$)**

| Parameter | $z_0$ | $z_{\pm1}$ | $z_{\pm2}$ | $z_{\pm3}$ | $z_{\pm4}$ | $z_{\pm5}$ |
|---|---|---|---|---|---|---|
| value | -16.49 | -0.04649 | -1.456 | 0.08187 | -0.08745 | -0.04282 |
| Parameter | $z_{\pm6}$ | $z_{\pm7}$ | $z_{\pm8}$ | $z_{\pm9}$ | $z_{\pm10}$ | $r_2$ |
| value | -0.1280 | -0.04414 | -0.06138 | 0.03527 | -0.09541 | $1.12\lambda_0$ |

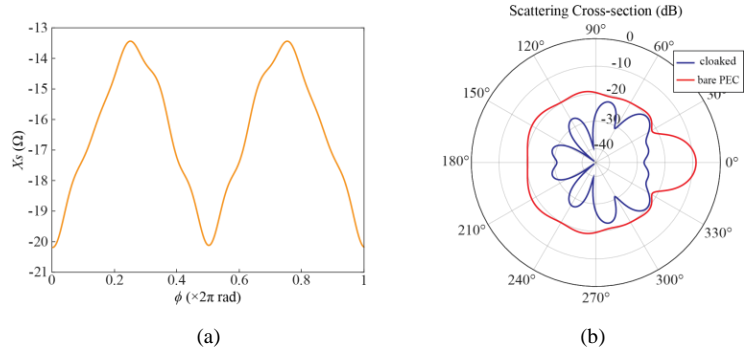

Fig. 4. (a) The optimized surface reactance value for cloaking under TM oblique incident plane wave. The impedance is isotropic along *z*-axis. (b) The scattering cross-sections of a cloaked conducting cylinder and the bare conducting core.

We also give the simulated total field for visual comparison in Fig. 5. The total fields in the presence of the cloaked conducting cylinder are listed on the first row while those in the presence of a bare conducting core are presented on the second row. We can observe clear "shadows" both in E-fields (*x*- and *z*-components) and H-field (*y*-component) due to the obstruction of the conducting cylinder. However, when the host medium together with the impedance surface are applied, the "shadow" is effectively compensated.

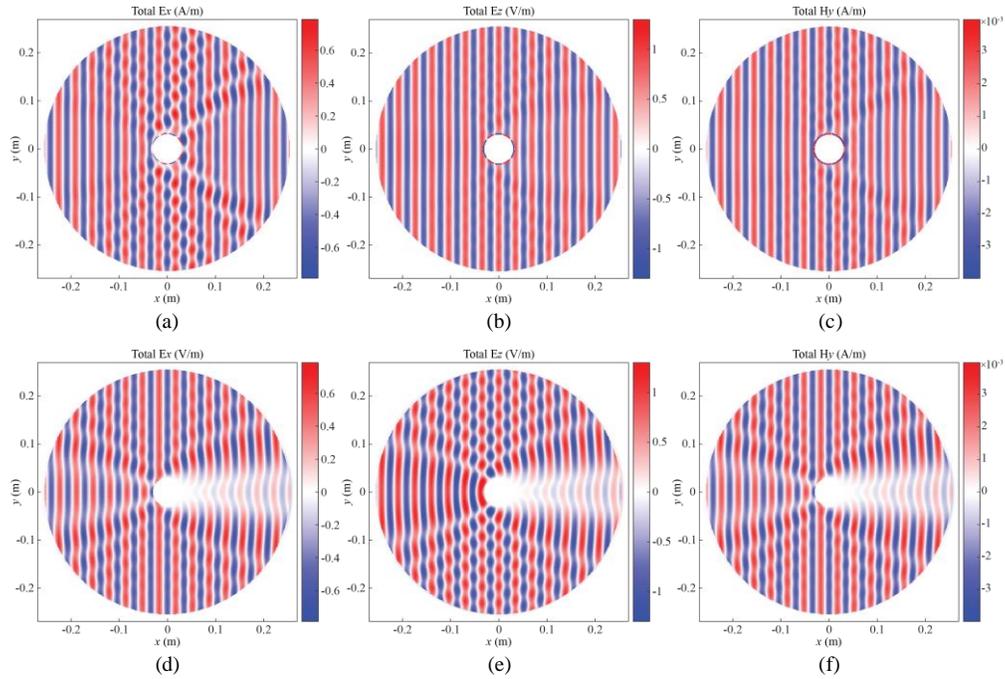

Fig. 5. The total field in the presence of a cloaked conducting cylinder (a) Ex field (b) Ez field and (c) Hy field. The total field in the presence of a bare conducting core (d) Ex field (e) Ez field and (f) Hy field.

[50, 51] provide a systematic approach to realize the surface impedance in physical way. We first design a metasurface with ideal surface impedance to cloak a cylinder under the normal illumination of TM wave. The radius of the conducting core is $\lambda_0$, and the relative permittivity

of the host medium is 16. Table 3 presents the optimized impedance Fourier coefficients and the outer radius of the host medium.

Table 3. The Optimizable Parameters in TM Normal Cloaking Scenario ($\theta_{in} = 90°$)

| Parameter | $z_0$ | $z_{\pm 1}$ | $z_{\pm 2}$ | $z_{\pm 3}$ | $z_{\pm 4}$ | $z_{\pm 5}$ |
|---|---|---|---|---|---|---|
| value | 20.81 | 0 | -2.068 | 0 | 0.1983 | 0 |
| Parameter | $z_{\pm 6}$ | $z_{\pm 7}$ | $z_{\pm 8}$ | $z_{\pm 9}$ | $z_{\pm 10}$ | $r_2$ |
| value | 0.08394 | 0 | 0.06955 | 0 | 0.05175 | 1.105 |

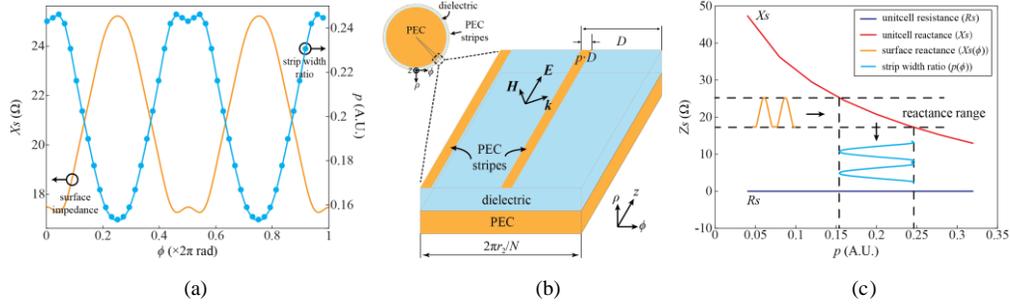

Fig. 6. (a) The optimized surface reactance for cloaking under TM normal incident wave (solid line) and corresponding strip width ratio ($p$) on each unit cell (dots). (b) Schematic of the unit cell. Each unit cell is a part of conducting-dielectric-metasurface structure. The whole surface is divided in to $N$ fractions, when $N$ is large or the curvature is small, each fraction can be treated as a planar structure. (c) The mapping between the surface reactance and the strip width ratio.

Fig. 6(a) shows that along $\phi$-dimension the reactance values ($Xs(\phi)$) are inductive. In the designing of the physical structure, the surface impedance is discretized into $N$ fractions ($N$ = 48 for this example). Each fraction of conducting-dielectric-metasurface structure consists of 2 grounded planar unit cells. The schematic of the fraction and unit cell are shown in Fig. 6(b). The unit cell is infinitely elongated in $z$-axis. On the surface of host medium, we use conducting stripe lines that parallel to electric field to synthesize inductive surface impedance [51]. In Fig. 6(b), $D$ denotes the spatial periodicity of the unit cell, and $p$ is the ratio between a single strip width and spatial periodicity $D$.

The mapping of the reactance value of the unit cell ($Xs$) and strip width ratio ($p$) is shown in Fig. 6(c). Since the reactance distribution (yellow solid line) is known, with this mapping, the strip width ratio ($p$) corresponding to the reactance can be obtained by mapping each reactance value ($Xs(\phi)$) to the mapping curve (red solid line), and then finding the $p(\phi)$ value on the horizontal axis. The blue dots in Fig. 6(a) are the strip width ratios for the conducting stripes in $N$ discretized fractions.

The results of total E-fields for TM normal illumination cloaking are shown in Fig. 7. In Fig. 7(a), the ideal surface impedance given in Table 3 is used. The cloaking performance is visibly satisfactory. For the uncloaked case shown in Fig. 7(b), the "shadow" of E$z$-field on the right side is obvious due to the blockage of the conducting cylinder. Then, the total E$z$-field in the presence cloaked cylinder with physically designed metasurface is presented in Fig. 7(c). Although the scattering level is visibly not as low as that in the presence of the cloaked cylinder with ideal metasurface, it still achieves on average -10dB SCS reduction compared to a bare conducting cylinder, as indicated in Fig. 7(e). The enlarged view of the E$z$-field near the physical metasurface is shown in Fig. 7(d). Note that the white "dots" on the interface indicating zero-E-field are caused by conducting stripes.

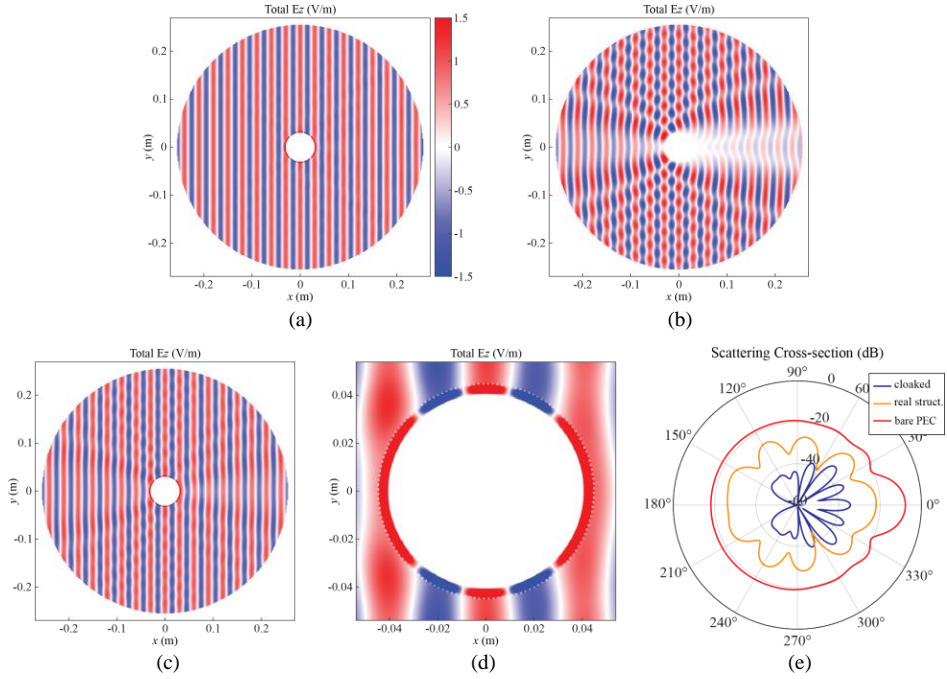

Fig. 7. Cloaking a cylinder under TM normal incident field. (a) The total E-field of the cloaked conducting cylinder using ideal impedance surface presented in Fig. 6(a). (b) The total E-field of the bare conducting core. (c) The total E-field of the cloaked conducting cylinder with physical impedance surface. (d) The enlarged view of total E-field adjacent to the conducting strip structure. The white "dots" on the interface indicating zero-E-field are caused by conducting stripes. (e) The scattering cross-sections of a cloaked conducting cylinder with ideal/physical impedance surface and the bare conducting core.

The proposed cloaking algorithm can also be applied to non-planar incident field cases, since any 2D incident field can be expanded with cylindrical harmonics whose coefficients are known. For a non-2D incident field case, when the cross section of the scatterer is small, the incident field can be locally approximated as 2D field near the scatterer, which requires the incident field and the matters distribution are locally invariant or harmonic in longitudinal dimension. To further illustrate this ability of our method, we present an example that cloaking a finite length cylinder under $z$-direction-oriented magnetic dipole illumination. The schematic is shown in Fig. 8(a). For convenience, the magnetic dipole is placed at the origin, and the radiation field of the magnetic dipole takes the form of

$$\begin{cases} \boldsymbol{E} = -\hat{a}_\varphi \frac{jkP_m}{4\pi}\sin\theta \frac{e^{-jkr}}{r} = \frac{jkP_m}{4\pi}\left(\hat{a}_x \frac{ye^{-jk\sqrt{x^2+y^2+z^2}}}{x^2+y^2+z^2} + \hat{a}_y \frac{xe^{-jk\sqrt{x^2+y^2+z^2}}}{x^2+y^2+z^2}\right) \\ \boldsymbol{H} = \hat{a}_\theta \frac{jkP_m}{4\pi\eta_0}\sin\theta \frac{e^{-jkr}}{r} \Rightarrow H_z = \frac{jkP_m}{4\pi\eta} \frac{x^2+y^2}{x^2+y^2+z^2} \frac{e^{-jk\sqrt{x^2+y^2+z^2}}}{\sqrt{x^2+y^2+z^2}} \end{cases}, \quad (18)$$

where $P_m$ is the magnetic dipole moment and $\eta_0$ is the wave impedance in free space. The axis of the cylinder is parallel to $z$-axis at $(x, y) = (6\lambda_0, 0)$. The radius of the conducting core is $0.5\lambda_0$, and the length of the cylinder is $4\lambda_0$ (from $z = -2\lambda_0$ to $z = 2\lambda_0$). The outer radius of the host medium is $1.117\lambda_0$ and the relative permittivity is 4. The cylinder can be piecewise sliced and perpendicular to the $z$-axis. Each slice of the cylinder is locally illuminated by a TE incident field (not a planar wave).

Based on our method, only the $z$-component of incident H-field is of our interest. It can be expanded by cylindrical harmonics. Note that the transverse wavenumbers $k_t$ used for cylindrical wave expansions are variant at different elevations. In the optimization process, we firstly obtain a solution for an arbitrary slice with incident angle $\theta_{inc}$ (denoted as Optim.1). This solution can be used as the initial optimization point for the adjacent slice with a little different incident angle $\theta_{inc}+\Delta\theta_{inc}$ (denoted as Optim.2). Although the optimization problem for our model is extremely non-linear, such strategy can facilitate the subsequent optimizations. Because negligible deviation of incident angle and difference of incident field between Optim.1 and Optim.2 is a small perturbation, a local minimum of Optim.1 is very near to a local minimum of Optim.2, so that the subsequent minimum searching becomes easier.

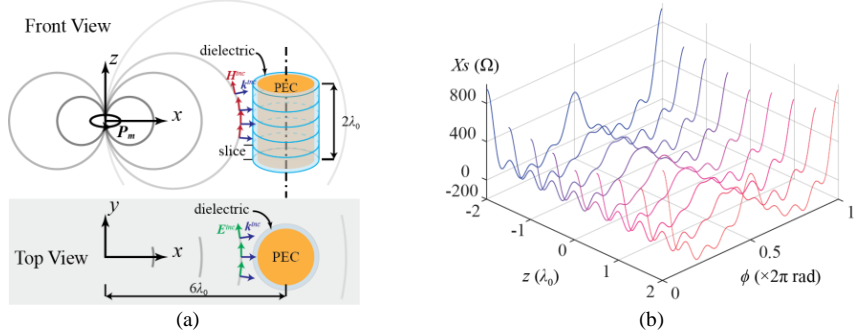

Fig. 8. (a) The schematic of a finite-length cylinder illuminated by a magnetic dipole field. The cylinder can be viewed as stacked slices. Each slice of the cylinder is locally illuminated by a non-planar TE incident field. (b) The optimized surface reactance for slices along $z$-axis.

Fig. 8(b) shows the optimized surface reactance. The value varies along both $\phi$- and $z$-dimensions, while the variation along $z$-dimension is very slow. This is beneficial for the approximation of local incident wave – the surface reactance can be approximated to be locally invariant in $z$-dimension.

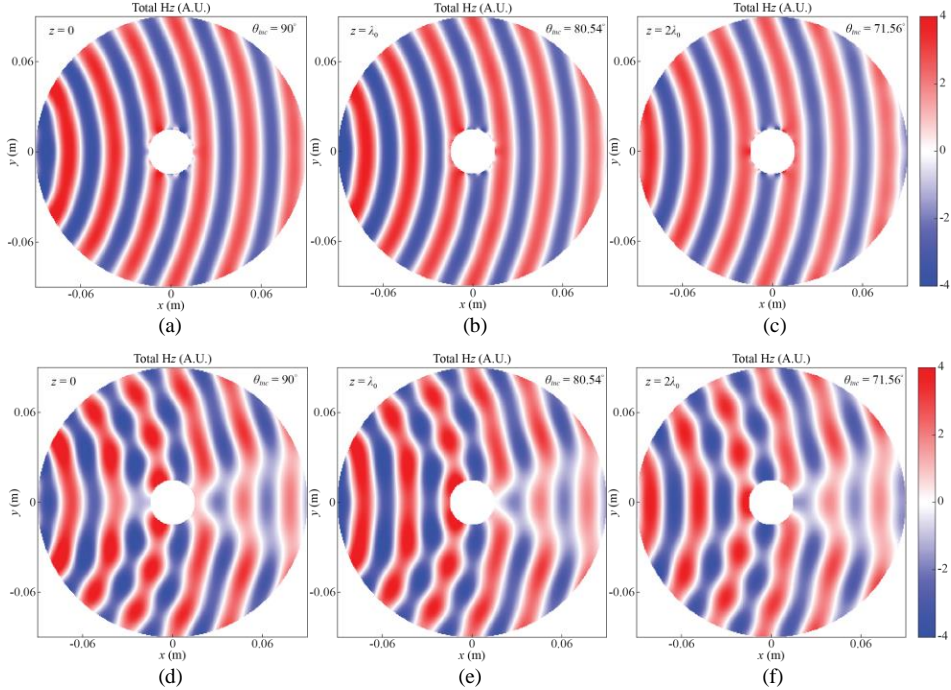

Fig. 9. The $z$-components of total field of magnetic field in the presence of a cloaked finite-length conducting cylinder at elevations (a) $z = 0$, corresponding incident angle $\theta_{inc} = 90°$ (b) $z = \lambda_0$, corresponding incident angle $\theta_{inc} = 80.54°$, and (c) $z = 2\lambda_0$, corresponding incident angle $\theta_{inc}$

= 71.56°. The z-components of total field of magnetic field in the presence of a bare conducting core at elevations (d) $z = 0$ (e) $z = \lambda_0$ and (f) $z = 2\lambda_0$.

Fig. 9(a)-(c) show the total $H_z$-fields near the cloaked conducting cylindrical core at the 3 different elevations, i.e., $z = 0$, $\lambda_0$ and $2\lambda_0$, respectively. It is obvious that the fields are maintained when illuminating on the finite-length cloaked cylinder. For comparison, the total $H_z$-fields at the same elevations in the presence of the bare conducting core are plotted in Fig. 9(d)-(f), in which the distortions of fields are visible.

## 4. Discussions and Conclusions

The above theoretical analysis and simulations show that this optimization method of anisotropic surface impedance is applicable for versatile 2D and quasi-3D cloaking. Besides, some extended mathematical and physical insights are worth discussing.

First, the truncation order. The cylindrical wave expansion format of the incident wave, e.g., plane wave, is the summation of infinite orders of cylindrical harmonics. It is impossible to take every order of harmonics into account. How to choose a proper highest order to truncate the summation? In practice, if the truncation of the summation is at the order $n = \pm l$, the result of the forward problem doesn't differ a lot from that with truncation order chosen at $n = \pm(l+1)$ or more, we say that the truncation order $l$ is high enough and the result is convergent with respect to the truncation order. One may think a much higher order can lead to a more accurate result. On the contrary, when an extremely high truncation order, the result becomes inaccurate due to the singularity of the matrix that includes Bessel functions of the first kind, since when $n$ goes higher, $J_n(k_0 r_2)$ approaches to 0.

Table 4. The Optimizable Parameters in TM Normal Cloaking Scenario ($\theta_{in} = 90°$) (An Alternative Solution)

| Parameter | $z_0$ | $z_{\pm 1}$ | $z_{\pm 2}$ | $z_{\pm 3}$ | $z_{\pm 4}$ | $z_{\pm 5}$ |
|---|---|---|---|---|---|---|
| value | 31.19 | 0 | -9.995 | 0 | 1.307 | 0.0002 |
| Parameter | $z_{\pm 6}$ | $z_{\pm 7}$ | $z_{\pm 8}$ | $z_{\pm 9}$ | $z_{\pm 10}$ | $r_2$ |
| value | 0.3674 | -0.0003 | 0.4776 | 0 | 0.2761 | 1.2144 |

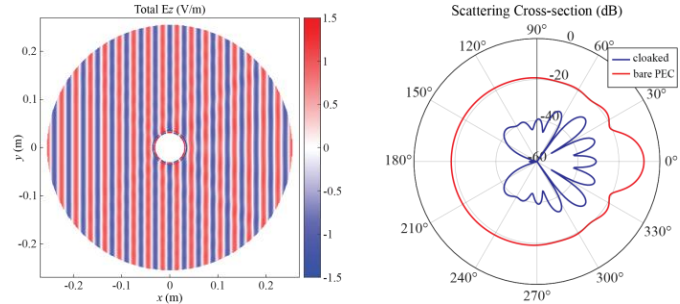

Fig. 10. An alternative solution for cloaking a cylinder under TM normal incident field. (a) The total E-field of the cloaked conducting cylinder. (b) The scattering cross-section of the cloaked conducting cylinder. The SCS of a bare conducting core is also plotted for comparison.

Second, the multiple solutions. For a given incident field and a given size of the conducting cylinder core, the proposed method renders multiple solution potential. In the optimization, the tunable parameters are continuous variables, such as Fourier coefficients of the surface impedance and the thickness of the host medium (dielectric), providing considerable freedoms. And the optimization problem itself is non-convex, with multiple local minima. pursuing the global minima of the problem is unnecessary because each local minimum, whose cost value is small enough, can be one effective implementation of cloaking. Furthermore, we can set multi-objective optimization problem to realize the cloaking for different incident field simultaneously, e.g., plane wave and cylindrical wave incident field. For illustration of multiple solution potential, we provide an alternative implementation other than Fig. 6 and Fig. 7 with the same radius of the conducting core ($r_1 = \lambda_0$) and the same permittivity ($\varepsilon_r = 16$) of the host medium the same incident field (TM normal incident field). Table 4 presents the optimizable parameters. The total E-field and the scattering cross-section is shown in Fig. 10.

Third, Omnidirectional cloaking ability. Omnidirectional cloaking is another application desire. In the previous works, homogeneous metamaterials or isotropic metasurface are adopted, which means $\phi$-invariant. In those cases, it is natural that the cloaking is omnidirectional. However, these works assume the size of the cylindrical or spherical scatterer is in quasi-static limit region (much less than a wavelength), so, only the heading orders of harmonics are considered. For a larger scatterer, it is necessary to cancel more orders of harmonics, in other word, the surface impedance should include more Fourier components to "bridge" more different orders of harmonics. Although it is difficult to realize omnidirectional cloaking, cloaking for a certain azimuth range, or cloaking for some discrete azimuth angle is possible by setting up multi-objective optimization problem.

In this work, we focus on the cloaking for a cylindrical conducting core under 2D and quasi-3D incident field. The field components are expressed in dependence of longitudinal component $H_z$ and $E_z$-field. The surface tangential E-field and surface equivalent electric current (tangential H-field difference across the surface) are linked by the surface impedance in Fourier series form. Optimizing the Fourier coefficients of the surface impedance and the thickness of the host medium, a low scattering performance can be achieved. With four full-wave simulation examples, we demonstrate that the proposed cloaking method is applicable for TE/TM normal/oblique planar/non-planar incident field and is adoptable for large-size cylindrical scatterers. The optimized surface impedance can be physically realized. This method can be further applied to terahertz and optical regimes.

**Funding**

China Scholarship Council

**Acknowledgement**

Y. Z., X. C. designed the research. All authors contributed to data interpretation and the composition of the manuscript.

**Disclosures**

The authors declare no conflicts of interest.

**Data Availability**

Data underlying the results presented in this paper are not publicly available at this time but may be obtained from the authors upon reasonable request.

**Supplemental document**

See Supplement 1 for supporting content.

**References**

1. Fleury, R. and A. Alu. Cloaking and invisibility: A review. in Forum for Electromagnetic Research Methods and Application Technologies (FERMAT). 2014.
2. Fleury, R., F. Monticone, and A. Alù, Invisibility and cloaking: Origins, present, and future perspectives. Physical Review Applied, 2015. **4**(3): p. 037001.
3. Pendry, J.B., D. Schurig, and D.R. Smith, Controlling electromagnetic fields. science, 2006. **312**(5781): p. 1780-1782.

4. Landy, N. and D.R. Smith, A full-parameter unidirectional metamaterial cloak for microwaves. Nature materials, 2013. **12**(1): p. 25-28.
5. Li, J. and J.B. Pendry, Hiding under the carpet: a new strategy for cloaking. Physical review letters, 2008. **101**(20): p. 203901.
6. Ma, H.F. and T.J. Cui, Three-dimensional broadband ground-plane cloak made of metamaterials. Nature communications, 2010. **1**(1): p. 21.
7. Liu, R., et al., Broadband ground-plane cloak. Science, 2009. **323**(5912): p. 366-369.
8. Chen, P.Y., J. Soric, and A. Alù, Invisibility and cloaking based on scattering cancellation. Advanced Materials, 2012. **24**(44): p. OP281-OP304.
9. Alù, A. and N. Engheta, Polarizabilities and effective parameters for collections of spherical nanoparticles formed by pairs of concentric double-negative, single-negative, and ∕ or double-positive metamaterial layers. Journal of Applied Physics, 2005. **97**(9).
10. Alù, A., Mantle cloak: Invisibility induced by a surface. physical review B, 2009. **80**(24): p. 245115.
11. Tricarico, S., F. Bilotti, and L. Vegni, Scattering cancellation by metamaterial cylindrical multilayers. Journal of the European Optical Society-Rapid publications, 2009. **4**.
12. Soric, J., et al., Radio-transparent dipole antenna based on a metasurface cloak. Nature communications, 2022. **13**(1): p. 1114.
13. Silveirinha, M.G., A. Alù, and N. Engheta, Parallel-plate metamaterials for cloaking structures. Physical Review E, 2007. **75**(3): p. 036603.
14. Edwards, B., et al., Experimental verification of plasmonic cloaking at microwave frequencies with metamaterials. Physical review letters, 2009. **103**(15): p. 153901.
15. Silveirinha, M.G., A. Alù, and N. Engheta, Infrared and optical invisibility cloak with plasmonic implants based on scattering cancellation. Physical Review B, 2008. **78**(7): p. 075107.
16. Danaeifar, M., N. Granpayeh, and M.R. Booket, Optical invisibility of cylindrical structures and homogeneity effect on scattering cancellation method. Electronics Letters, 2016. **52**(1): p. 29-31.
17. Song, Z., et al., Making a continuous metal film transparent via scattering cancellations. Applied Physics Letters, 2012. **101**(18).
18. Tretyakov, S., et al., Broadband electromagnetic cloaking of long cylindrical objects. Physical review letters, 2009. **103**(10): p. 103905.
19. Silveirinha, M.G., A. Alù, and N. Engheta, Cloaking mechanism with antiphase plasmonic satellites. Physical Review B, 2008. **78**(20): p. 205109.
20. Farhat, M., et al., Scattering cancellation of the magnetic dipole field from macroscopic spheres. Optics express, 2012. **20**(13): p. 13896-13906.
21. Fruhnert, M., et al., Tunable scattering cancellation cloak with plasmonic ellipsoids in the visible. Physical Review B, 2016. **93**(24): p. 245127.
22. Li, T., et al., In-band Radar Cross Section reduction for electromagnetic window by simultaneously enhancing transmission and coding reflection. Journal of Physics D: Applied Physics, 2022. **55**(49): p. 49LT01.
23. Ji, C., et al., Broadband low-scattering metasurface using a combination of phase cancellation and absorption mechanisms. Optics Express, 2019. **27**(16): p. 23368-23377.
24. Zheng, Y., et al., Wideband RCS reduction of a microstrip antenna using artificial magnetic conductor structures. IEEE antennas and wireless propagation letters, 2015. **14**: p. 1582-1585.
25. Yu, N., et al., Light propagation with phase discontinuities: generalized laws of reflection and refraction. science, 2011. **334**(6054): p. 333-337.
26. Ni, X., et al., An ultrathin invisibility skin cloak for visible light. Science, 2015. **349**(6254): p. 1310-1314.
27. Yang, Y., et al., Full‐polarization 3D metasurface cloak with preserved amplitude and phase. Advanced Materials, 2016. **28**(32): p. 6866-6871.
28. Shi, S., et al., Oblique incident achromatic cloaking based on all-dielectric multilayer frame metasurfaces. Applied Physics A, 2022. **128**(12): p. 1085.
29. Pestourie, R., et al., Inverse design of large-area metasurfaces. Optics express, 2018. **26**(26): p. 33732-33747.
30. Kwon, D.-H. and S.A. Tretyakov, Perfect reflection control for impenetrable surfaces using surface waves of orthogonal polarization. Physical Review B, 2017. **96**(8): p. 085438.
31. Kwon, D.-H., Modulated reactance surfaces for leaky-wave radiation based on complete aperture field synthesis. IEEE Transactions on Antennas and Propagation, 2020. **68**(7): p. 5463-5477.
32. Budhu, J., L. Szymanski, and A. Grbic, Design of planar and conformal, passive, lossless metasurfaces that beamform. IEEE Journal of Microwaves, 2022. **2**(3): p. 401-418.
33. Budhu, J., L. Szymanski, and A. Grbic, Design of passive and lossless single layer metasurfaces for far field beamforming. arXiv preprint arXiv:2112.03250, 2021. **2**(3): p. 401-418.
34. Kuester, E.F., et al., Averaged transition conditions for electromagnetic fields at a metafilm. IEEE Transactions on Antennas and Propagation, 2003. **51**(10): p. 2641-2651.
35. Pfeiffer, C. and A. Grbic, Metamaterial Huygens' surfaces: tailoring wave fronts with reflectionless sheets. Physical review letters, 2013. **110**(19): p. 197401.
36. Campione, S., et al., Tailoring dielectric resonator geometries for directional scattering and Huygens' metasurfaces. Optics Express, 2015. **23**(3): p. 2293-2307.
37. Kim, M., A.M. Wong, and G.V. Eleftheriades, Optical Huygens' metasurfaces with independent control of the magnitude and phase of the local reflection coefficients. Physical Review X, 2014. **4**(4): p. 041042.

38. Gao, X., et al., A reconfigurable broadband polarization converter based on an active metasurface. IEEE Transactions on Antennas and Propagation, 2018. **66**(11): p. 6086-6095.
39. Selvanayagam, M. and G.V. Eleftheriades, An active electromagnetic cloak using the equivalence principle. IEEE Antennas and Wireless Propagation Letters, 2012. **11**: p. 1226-1229.
40. Selvanayagam, M. and G.V. Eleftheriades, Experimental demonstration of active electromagnetic cloaking. Physical review X, 2013. **3**(4): p. 041011.
41. Sounas, D.L., R. Fleury, and A. Alù, Unidirectional cloaking based on metasurfaces with balanced loss and gain. Physical Review Applied, 2015. **4**(1): p. 014005.
42. Lee, H. and D.-H. Kwon, Microwave metasurface cloaking for freestanding objects. Physical Review Applied, 2022. **17**(5): p. 054012.
43. Kwon, D.-H., Lossless tensor surface electromagnetic cloaking for large objects in free space. Physical Review B, 2018. **98**(12): p. 125137.
44. Kwon, D.-H., Illusion electromagnetics for free-standing objects using passive lossless metasurfaces. Physical Review B, 2020. **101**(23): p. 235135.
45. Teperik, T.V. and A. de Lustrac, Electromagnetic cloak to restore the antenna radiation patterns affected by nearby scatter. AIP Advances, 2015. **5**(12).
46. Dehmollaian, M., G. Lavigne, and C. Caloz, Transmittable Nonreciprocal Cloaking. Physical Review Applied, 2023. **19**(1): p. 014051.
47. Hamzavi-Zarghani, Z., A. Yahaghi, and L. Matekovits, Electrically tunable mantle cloaking utilizing graphene metasurface for oblique incidence. AEU-International Journal of Electronics and Communications, 2020. **116**: p. 153080.
48. Balanis, C.A., Advanced engineering electromagnetics. 2012: John Wiley & Sons.
49. Wait, J.R., Scattering of a plane wave from a circular dielectric cylinder at oblique incidence. Canadian journal of physics, 1955. **33**(5): p. 189-195.
50. Wang, X.-c., et al., Systematic design of printable metasurfaces: Validation through reverse-offset printed millimeter-wave absorbers. IEEE Transactions on Antennas and Propagation, 2018. **66**(3): p. 1340-1351.
51. Luukkonen, O., et al., Simple and accurate analytical model of planar grids and high-impedance surfaces comprising metal strips or patches. IEEE Transactions on Antennas and Propagation, 2008. **56**(6): p. 1624-1632.